\documentclass[aps,pra,amsmath,amssymb,tightenlines,epsfig,floatfix,superscriptaddress, twocolumn, longbibliography, standalone]{revtex4-1}
\usepackage{graphicx}
\usepackage{dcolumn}	
\usepackage{bm}			
\usepackage{amsfonts}
\usepackage{xspace}
\usepackage{color}
\usepackage{epstopdf}
\usepackage{multirow}
\usepackage{braket}

\begin{document}

\newcommand{\psihat}{\ensuremath{\hat{\psi}}\xspace}
\newcommand{\psihatd}{\ensuremath{\hat{\psi}^{\dagger}}\xspace}
\newcommand{\ahat}{\ensuremath{\hat{a}}\xspace}
\newcommand{\Ham}{\ensuremath{\mathcal{H}}\xspace}
\newcommand{\ahatd}{\ensuremath{\hat{a}^{\dagger}}\xspace}
\newcommand{\bhat}{\ensuremath{\hat{b}}\xspace}
\newcommand{\bhatd}{\ensuremath{\hat{b}^{\dagger}}\xspace}
\newcommand{\boldr}{\ensuremath{\mathbf{r}}\xspace}
\newcommand{\dr}{\ensuremath{\,d^3\mathbf{r}}\xspace}
\newcommand{\dy}{\ensuremath{\,d\mathbf{y}}\xspace}
\newcommand{\dz}{\ensuremath{\,d\mathbf{z}}\xspace}
\newcommand{\tr}{\ensuremath{\,\mathrm{Tr}}\xspace}
\newcommand{\dk}{\ensuremath{\,d^3\mathbf{k}}\xspace}
\newcommand{\etal}{\emph{et al.\/}\xspace}
\newcommand{\ie}{i.e.\ }
\newcommand{\eq}[1]{Eq.\,(\ref{#1})\xspace}
\newcommand{\fig}[1]{Fig.\,\ref{#1}\xspace}
\newcommand{\abs}[1]{\left| #1 \right|}
\newcommand{\proj}[2]{\left| #1 \rangle\langle #2\right| \xspace}
\newcommand{\Qhat}{\ensuremath{\hat{Q}}\xspace}
\newcommand{\Qhatd}{\ensuremath{\hat{Q}^\dag}\xspace}
\newcommand{\phihatd}{\ensuremath{\hat{\phi}^{\dagger}}\xspace}
\newcommand{\phihat}{\ensuremath{\hat{\phi}}\xspace}
\newcommand{\boldk}{\ensuremath{\mathbf{k}}\xspace}
\newcommand{\boldp}{\ensuremath{\mathbf{p}}\xspace}
\newcommand{\boldsigma}{\ensuremath{\boldsymbol\sigma}\xspace}
\newcommand{\boldalpha}{\ensuremath{\boldsymbol\alpha}\xspace}
\newcommand{\grad}{\ensuremath{\boldsymbol\nabla}\xspace}
\newcommand{\parti}[2]{\frac{ \partial #1}{\partial #2} \xspace}
 \newcommand{\vs}[1]{\ensuremath{\boldsymbol{#1}}\xspace}
\renewcommand{\v}[1]{\ensuremath{\mathbf{#1}}\xspace}
\newcommand{\Psihat}{\ensuremath{\hat{\Psi}}\xspace}
\newcommand{\Psihatd}{\ensuremath{\hat{\Psi}^{\dagger}}\xspace}
\newcommand{\Vhatd}{\ensuremath{\hat{V}^{\dagger}}\xspace}
\newcommand{\Xhat}{\ensuremath{\hat{X}}\xspace}
\newcommand{\Xhatd}{\ensuremath{\hat{X}^{\dag}}\xspace}
\newcommand{\Yhat}{\ensuremath{\hat{Y}}\xspace}
\newcommand{\Jhat}{\ensuremath{\hat{J}}\xspace}
\newcommand{\Yhatd}{\ensuremath{\hat{Y}^{\dag}}\xspace}
\newcommand{\Uhat}{\ensuremath{\hat{U}^{\dag}}\xspace}
\newcommand{\jhat}{\ensuremath{\hat{J}}\xspace}
\newcommand{\lhat}{\ensuremath{\hat{L}}\xspace}
\newcommand{\Nhat}{\ensuremath{\hat{N}}\xspace}
\newcommand{\rhohat}{\ensuremath{\hat{\rho}}\xspace}
\newcommand{\ddt}{\ensuremath{\frac{d}{dt}}\xspace}
\newcommand{\nset}{\ensuremath{n_1, n_2,\dots, n_k}\xspace}
\newcommand{\Var}{\ensuremath{\mathrm{Var}}\xspace}
\newcommand{\Erf}{\ensuremath{\mathrm{Erf}}\xspace}
\newcommand{\tprep}{\ensuremath{\tau_\text{prep}}\xspace}
\newcommand{\tint}{\ensuremath{\tau_\text{int}}\xspace}
\newcommand{\sigmahat}{\ensuremath{\hat{\sigma}}\xspace}
\newcommand{\phat}{\ensuremath{\hat{p}}\xspace}
\newcommand{\xhat}{\ensuremath{\hat{x}}\xspace}
\newcommand{\zhat}{\ensuremath{\hat{z}}\xspace}

\newcommand{\notes}[1]{{\color{blue}#1}}
\newcommand{\sah}[1]{{\color{magenta}#1}}
\newcommand{\Junbang}[1]{{\color{red}#1}}

\title{Twist and Turn Squeezing in a Multi-Mode Bose-Einstein Condensate}
\author{Junbang Liu} 
\affiliation{Department of Quantum Science, Research School of Physics, Australian National University, Canberra, Ngunnawal Country, Australia.}
\author{Thomas Bartlett}
\affiliation{Department of Quantum Science, Research School of Physics, Australian National University, Canberra, Ngunnawal Country, Australia.}
\author{Joseph Hope}
\affiliation{Department of Quantum Science, Research School of Physics, Australian National University, Canberra, Ngunnawal Country, Australia.}
\author{Simon Haine}
\affiliation{Department of Quantum Science, Research School of Physics, Australian National University, Canberra, Ngunnawal Country, Australia.}
\email{simon.a.haine@gmail.com}

\begin{abstract}
Here we examine the generation of Twist and Turn (TNT) Squeezing in a large atom-number Bose-Einstein Condensate for the purposes of generating quantum-enhanced states for atom interferometry. Unlike previous analysis, we examine situations where the multi-mode dynamics is significant, and cannot be captured by a simple single-mode model.  We find that in some regimes, with careful choice of the rotation parameter, we can still obtain squeezing much more rapidly than via one-axis twisting (OAT). 
\end{abstract}

\maketitle

\section{Introduction} \label{sec1}
\noindent There is currently considerable interest in the production of entangled states in Bose-Einstein condensates with the motivation of enhancing the sensitivity of atom interferometers and atomic clocks \cite{Pezze_review:2018, Szigeti:2021}. Without many-particle entanglement, the phase sensitivity of such experiments is fundamentally constrained to the shot-noise limit (SNL) $\Delta \phi =1/\sqrt{N}$ \cite{Giovannetti:2006, Pezze:2009}. In recent years, experiments in atomic systems based on the one-axis twisting (OAT) squeezing scheme of Kitagawa and Ueda \cite{Kitagawa:1993, Molmer:1999} have demonstrated metrologically useful spin-squeezing \cite{Esteve:2008, Leroux:2010}, and sub-shot-noise phase detection \cite{Gross:2010, Riedel:2010, Muessel:2014} in proof-of-principle experiments. However, typical experiments are limited to only moderate quantum enhancement due to constraints on the state preparation time imposed by dephasing \cite{Li:2008, Li:2009, Haine:2018}, multi-mode dynamics \cite{Haine:2009, Haine:2014}, or in the case of atomic gravimetry, expansion of the freely propagating wave-packets \cite{Szigeti:2020}. This leads to a degree of quantum enhancement that is considerably less than the theoretical optimum. Recently, a related method known as Twist and Turn (TNT) squeezing \cite{Micheli:2003, Sorelli:2019, Mirkhalaf:2018} has been demonstrated \cite{Strobel:2014, Muessel:2015}. The TNT Hamiltonian, which is a specific case of the Lipkin-Meshkov-Glick Hamiltonian \cite{Lipkin:1965} using the same nonlinear interactions as OAT with an additional linear rotation, typically reaches larger degrees of quantum enhancement for the same interaction time. As TNT is based on the same interactions that leads to OAT squeezing, it can be implemented in the same experimental set-ups. 

As well as spin-squeezing, it has been shown that TNT dynamics is capable of generating highly entangled states beyond the Gaussian regime \cite{Strobel:2014, Muessel:2015}. In this case, the metrologically useful entanglement can be quantified via the quantum Fisher information (QFI) \cite{Demkowicz-Dobrzanski:2014, Braunstein:1994, Paris:2009, Toth:2014}, and can be accessed via interaction based readouts \cite{Davis:2016, Hosten:2016b, Frowis:2016, Macri:2016, Linnemann:2016, Nolan:2017b, Szigeti:2017, Anders:2018, Mirkhalaf:2018, Haine:2018b, Hayes:2018, Haine:2021}. 

So far, experimental demonstrations of TNT dynamics have been restricted to samples of a few hundred atoms. One reason for this is that larger atom numbers introduce multimode dynamics \cite{Li:2009, Haine:2009, Opanchuk:2012, Haine:2014}. These dynamics can cause the two spin components to spatially separate. In the case of OAT, this separation can significantly increase the rate of entangling dynamics \cite{Riedel:2010, Li:2008, Nolan:2018}. However, for efficient TNT dynamics, we require continuous spatial overlap to enable continuous coupling between the two spin components. In this paper, we investigate how this multimode dynamics effects the implementation of TNT dynamics. 

The structure of this paper is as follows: In section \ref{sec2} we introduce the theoretical model used. In section \ref{sec3} we derive an effective single-mode model from our multi-mode model, and recap ideal single-mode behaviour of both the OAT and TNT hamiltonians. In section \ref{sec4} we identify three parameter regimes, and discuss how multi-mode dynamics effects the ability to generate spin-squeezing and entanglement in each of these regimes, before summarising our findings in section \ref{sec5}.

\section{Model} \label{sec2}
We consider an atomic BEC with two relevant atomic states $|a\rangle$ and $|b\rangle$, confined in a potential $V(x)$. For example, these levels could be the two hyperfine ground states of $^{87}$Rb, in which case $|a\rangle \equiv |F=1, m_F =0\rangle$ and $|b\rangle \equiv |F=2, m_F=0\rangle$. We also allow for continuous coupling between the two states via a radio frequency or microwave transition, which provides the `turn' in the TNT dynamics. Introducing the bosonic field operators $\psihat_i(\boldr)$, which annihilate an atom of state $|i\rangle$ from position $\boldr$, and obey the usual bosonic commutation relations
\begin{align}
    \left[\psihat_i(\boldr) \, , \, \, \psihatd_j(\boldr^\prime)\right] = \delta_{ij}\delta(\boldr-\boldr^\prime) \, ,
\end{align}
the Hamiltonian describing the system is
\begin{align}
\hat{H} &= \sum_{j=a,b} \int \psihatd_j(\boldr)\hat{H}_0 \psihat_j(\boldr) \, \dr  \nonumber\\
&+ \sum_{j,k = a,b}\frac{U_{ij}}{2} \int \psihatd_j(\boldr)\psihatd_k(\boldr)\psihat_j(\boldr)\psihat_k(\boldr)\, \dr \nonumber \\
&+ \frac{\hbar\Omega}{2}\int \left(\psihatd_a(\boldr)\psihat_b(\boldr) + \psihatd_b(\boldr)\psihat_a(\boldr)\right) \, \dr \, , \label{Ham_full}
\end{align}
where
\begin{align}
\hat{H}_0 &= \frac{-\hbar^2}{2m}\grad^2 + V(x)
\end{align}
is the single particle Hamiltonian, and 
\begin{equation}
U_{ij} = \frac{4\pi \hbar^2a_{ij}}{m}    
\end{equation}
is the inter-particle interactions between atoms in states $|i\rangle$ and $|j\rangle$, parameterised by the $s$-wave scattering length $a_{ij}$ \cite{Chin:2010}. The final term in \eq{Ham_full} describes continuous coupling between states $|a\rangle$ and $|b\rangle$ with Rabi frequency $\Omega$. We assume that all atoms are initially in state $|a\rangle$ in the ground motional state before applying a $\frac{\pi}{2}$ pulse at $t=0$ to coherently couple 50\% of the population to state $|b\rangle$, such that the state of each atom is $\frac{1}{\sqrt{2}}\left(|a\rangle + |b\rangle \right)$. The system then evolves under \eq{Ham_full}.

The goal of the scheme is to produce an entangled state, which when used as the input state of an atom interferometer, is capable of providing sensitivities better than the SNL. Atom interferometery is best described via the SU(2) collective pseudo-spin operators defined by
\begin{subequations}
\begin{align}
\hat{J}_x &= \frac{1}{2} \int  \left(\psihatd_b(\boldr)\psihat_a(\boldr) + \psihatd_a(\boldr)\psihat_b(\boldr)\right) \dr  \, , \\
\hat{J}_y &= - \frac{i}{2} \int \left(\psihatd_b(\boldr)\psihat_a(\boldr) - \psihatd_a(\boldr)\psihat_b(\boldr)\right) \dr \, , \\
\hat{J}_z &= \frac{1}{2} \int \left(\psihatd_a(\boldr)\psihat_a(\boldr) - \psihatd_b(\boldr)\psihat_b(\boldr)\right) \dr \, .
\end{align}
\label{Jdefine}
\end{subequations}

Assuming a simple Mach-Zehnder interferometer composed of a $\frac{\pi}{2}-\pi-\frac{\pi}{2}$ pulse sequence, if measurements capable of resolving the mean and variance of the number difference at the output of the interferometer are made, the phase sensitivity is given by
\begin{align}
    \Delta \phi &=\frac{\xi}{\sqrt{N}} \, ,
\end{align}
where $\xi$ is the Wineland spin-squeezing parameter \cite{Wineland:1992} defined by
\begin{align}
\xi &= \sqrt{\frac{N \mathrm{Var}(\jhat_{\theta})}{\langle \jhat_x\rangle^2}} \, ,
\end{align}
with
\begin{align}
    \jhat_\theta &= \jhat_z \cos \theta + \jhat_y \sin \theta \, ,
\end{align}
where the angle $\theta$ is chosen to minimise $\mathrm{Var}(\jhat_\theta)$. This state is then converted into a state with reduced fluctuations in number difference by a rotation by $-\theta$ around the $\hat{J}_x$, before entering the interferometer. In practice, this is done via an additional coherent Rabi pulse.

\section{Ideal single-mode dynamics} \label{sec3}
We begin by revising the ideal \emph{single-mode} dynamics of both the OAT and TNT Hamiltonians. We do this by expanding the field operator into an orthonormal single-particle basis, and then make the assumption that only one mode is occupied, ie
\begin{subequations}
\begin{align}
    \psihat_a(\boldr) &= \sum_{j=0}^\infty \ahat_j u_{a,j}(\boldr)  \approx u_{a,0}(\boldr) \ahat_0 \\
    \psihat_b(\boldr) &= \sum_{j=0}^\infty \bhat_j u_{b,j}(\boldr)  \approx u_{b,0}(\boldr) \bhat_0 \,. 
\end{align}
\end{subequations}
This approximation has been shown to be reasonably valid in tight confining potentials with small numbers of atoms \cite{Gross:2010, Riedel:2010}, but breaks down when the population increases to the point where the energy due to atomic interactions dominate \cite{Haine:2014}. Making this approximation in \eq{Ham_full} gives
\begin{align}
    \hat{H} &= \hbar\chi_{aa}\ahatd\ahatd\ahat\ahat + \hbar\chi_{bb}\bhatd\bhatd\bhat\bhat + 2\hbar\chi_{ab}\ahatd\ahat\bhatd\bhat  \nonumber \\
    &+ \frac{\hbar\Omega}{2}\left(\eta \ahatd \bhat + \eta^*\bhatd \ahat\right) \, , \label{Ham_singlemode}
\end{align}
where 
\begin{align}
    \chi_{ij} = \frac{U_{ij}}{2\hbar}\int \abs{u_{i}(\boldr)}^2 \abs{u_{j}(\boldr)}^2 \dr \, ,  \label{chi_est}
\end{align}
and
\begin{align}
\eta = \int u_a^*(\boldr) u_b(\boldr) \dr \, ,
\end{align}
and we have made the replacements $\ahat_0 \rightarrow \ahat$, $\bhat_0\rightarrow \bhat$, $u_{a(b),0}(\boldr) \rightarrow u_{a(b)}(\boldr)$ for notational simplicity. Introducing the single-mode version of the pseudo-spin operators (\eq{Jdefine})
\begin{subequations}
\begin{align}
    \Jhat_x &= \frac{1}{2}\left(\ahatd\bhat + \bhatd\ahat\right) \, , \\
    \Jhat_y &= \frac{i}{2}\left(\ahatd\bhat - \bhatd\ahat\right) \, , \\
    \Jhat_z &= \frac{1}{2}\left(\ahatd\ahat - \bhatd\bhat\right) \, ,
\end{align}
\end{subequations}
and assuming perfect overlap ($\eta =1$), the Hamiltonian (\eq{Ham_singlemode}) can be written as
\begin{align}
    \hat{H} &= \hbar \chi \Jhat_z^2 +\hbar\chi_- (\hat{N}-1)\Jhat_z + \hbar\Omega \Jhat_x \label{Ham_J} \, ,
\end{align}
where $\chi = \chi_{aa} + \chi_{bb} - 2\chi_{ab}$, $\chi_- = \chi_{aa}-\chi_{bb}$, and $\hat{N} = \ahatd\ahat + \bhatd\bhat$ is the total number of atoms.  For a fixed number of atoms $N$, when $\chi_- =0$, and $\Omega = \chi N/2$, this is the well studied TNT Hamiltonian, and when $\Omega=0$ we recover the OAT Hamiltonian. Specifically, we define
\begin{subequations}
\begin{align}
    \hat{H}_\mathrm{OAT} &= \hbar \chi \Jhat_z^2 \label{Ham_OAT} \, , \\ 
    \hat{H}_\mathrm{TNT} &= \hbar \chi \Jhat_z^2 +\Omega \Jhat_x \, . \label{Ham_TNT}
\end{align}
\end{subequations}
In writing \eq{Ham_OAT} and \eq{Ham_TNT}, we have neglected terms that only depend on $\hat{N}$, as they do not result in any observable effects.

\begin{figure}
    \centering
    \includegraphics[width=\columnwidth]{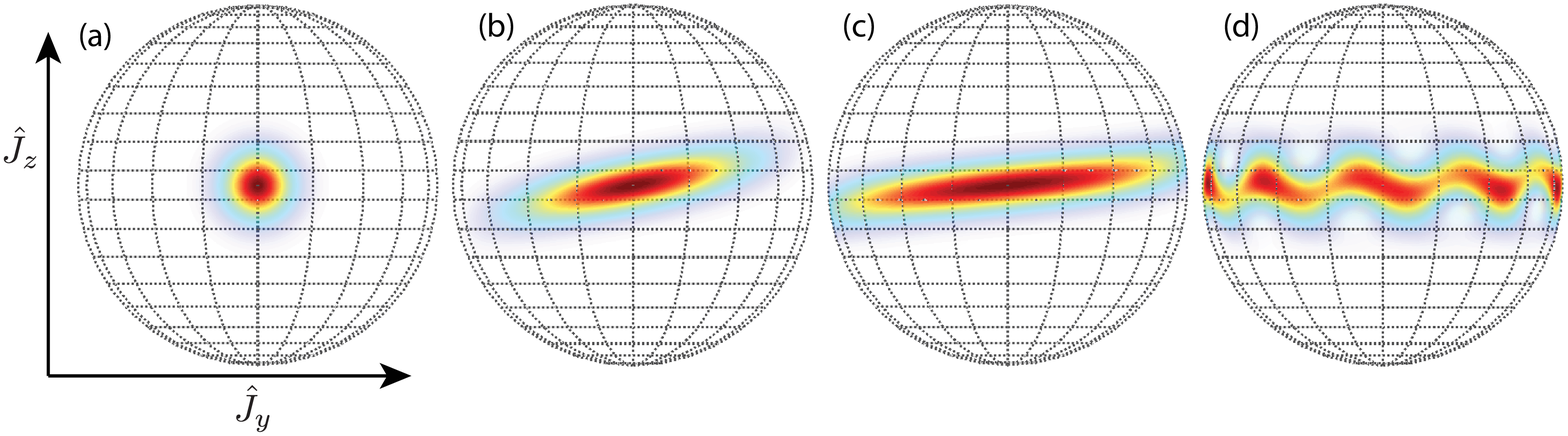}
    \caption{The Husimi $Q$-function $Q(\theta, \varphi)$, for the state evolving under \eq{Ham_OAT} for $N = 100$. The $Q$-function is defined as $Q(\theta, \varphi) = \abs{\langle\theta,\varphi|\Psi(t)\rangle}^2$, where $|\theta,\varphi\rangle = e^{i \varphi \hat{J}_z}e^{i \theta \hat{J}_y} |J_z = N/2\rangle$ represents the spin coherent 
state along $\theta$ and $\varphi$ directions corresponding to rotating the maximal $\hat{J}_z$ eigenstate around azimuthal  and polar angles $\{\theta, \varphi\}$ \cite{Arecchi:1972}. (a): $\chi t = 0$ and (b): $\chi t = 0.05$, (c): $\chi t = 0.1$, (d): $\chi t = 0.3$}
\label{fig1}
\end{figure}

\begin{figure}
    \centering
    \includegraphics[width=\columnwidth]{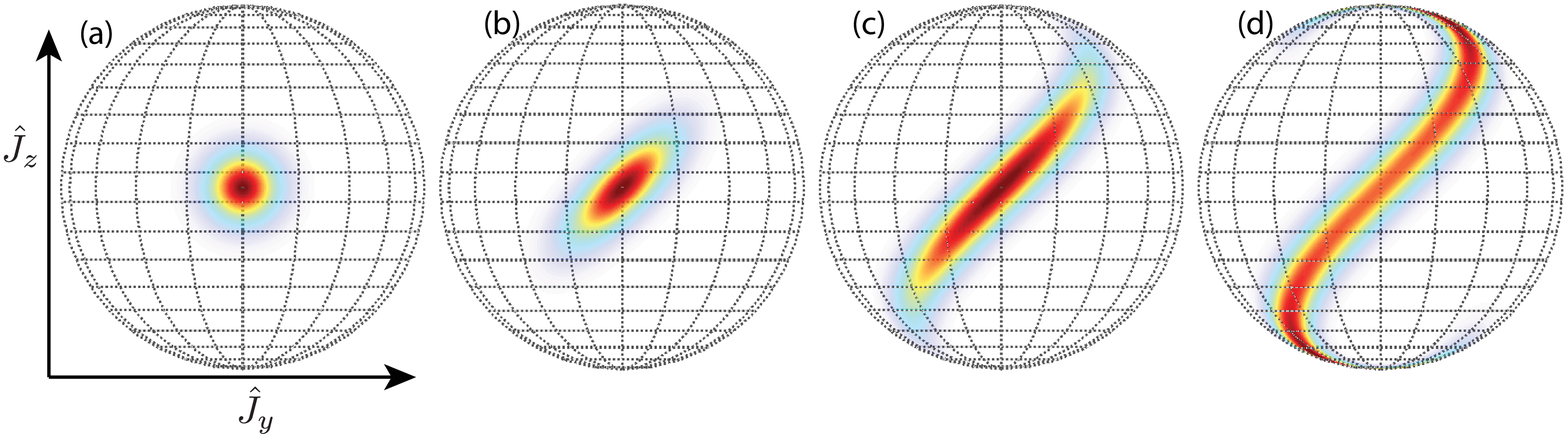}
    \caption{The Husimi $Q$-function $Q(\theta, \varphi)$, for the state evolving under \eq{Ham_TNT} for $N = 100$. The $Q$-function is defined as $Q(\theta, \varphi) = \abs{\langle\theta,\varphi|\Psi(t)\rangle}^2$, where $|\theta,\varphi\rangle = e^{i \varphi \hat{J}_z}e^{i \theta \hat{J}_y} |J_z = N/2\rangle$ represents the spin coherent 
state along $\theta$ and $\varphi$ directions corresponding to rotating the maximal $\hat{J}_z$ eigenstate around azimuthal  and polar angles $\{\theta, \varphi\}$ \cite{Arecchi:1972}. (a): $\chi t = 0$ and (b): $\chi t = 0.02$, (c): $\chi t = 0.04$, (d): $\chi t = 0.06$}
\label{fig2}
\end{figure}

\begin{figure}[htbp]
\includegraphics[width=\columnwidth]{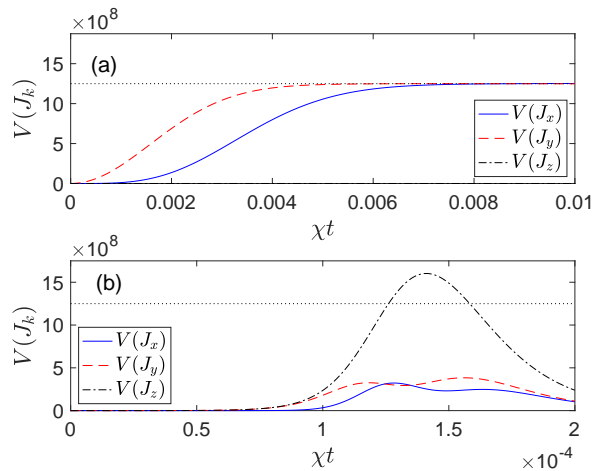}
\caption{Variance of the three components of the pseudo-spin for (a) OAT and (b) TNT dynamics. The black dotted line represents the threshold $N^2/8$, which is the approximate value that OAT dynamics plateaus at for long time. The number of atoms was $N = 10^5$.
}
\label{fig:fig3}
\end{figure}

Figures \ref{fig1} and \ref{fig2} show the evolution of an initial coherent spin state (CSS) under these Hamiltonians. We visualise the state using the Husimi $Q$-function \cite{Arecchi:1972, Agarwal:1998}. The most notable difference between OAT and TNT dynamics, is the speed at which the state evolves away from a CSS. Figure (\ref{fig:fig3}) shows the variance of the three pseuedo-spins ($\Jhat_x$, $\Jhat_y$, $\Jhat_z$) for both cases. We see that the timescale for the variance to increase from $\sim N/4$ (that of a CSS) to $\sim N^2/8$ (that of a highly non-classical state, such as a twin Fock-state, for example) is approximately 40 times faster for the TNT dynamics.

\begin{figure}[htbp]
\includegraphics[width=\columnwidth]{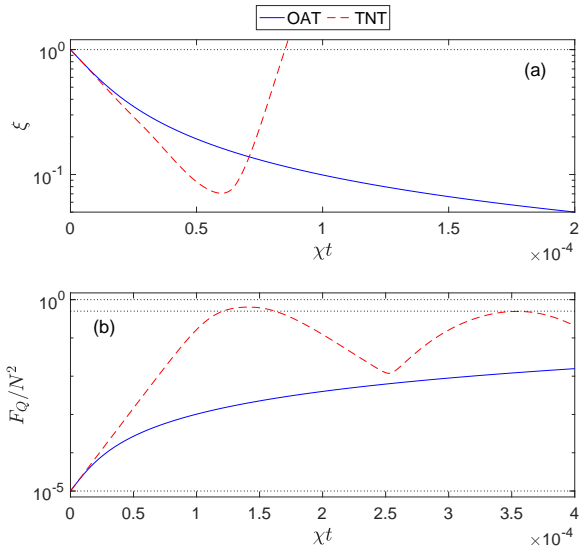}
\caption{(a) Evolution of the spin squeezing parameter for OAT (blue solid line) and TNT dynamics (red dashed line). (b) The QFI for OAT (blue solid line) and TNT (red dashed line). The three horizontal black dotted lines represent $F_Q = N$ (the limit for unentangled particles), $N^2/2$, (the long-time plateu for OAT dynamics)  and $N^2$ (the Heisenberg limit). The number of particles was $N=10^5$.}
\label{fig4}
\end{figure}

Two quantities of significant interest for sensing applications are the spin-squeezing parameter $\xi$, and the quantum Fisher information. Figure (\ref{fig4}a) shows the spin-squeezing parameter for both cases. We see that in the case of TNT, $\xi$ decreases significantly faster than for OAT, and OAT reaches a lower minimum. Specifically, it takes OAT more than twice as long to reach the minimum $\xi$ from TNT dynamics. Alternatively, if the state preparation time is constrained, TNT achieves a value of $\xi$ 2.3 times smaller for the same state preparation time. Beyond this minimum, `\emph{oversqueezing}' prevents any further reduction in $\xi$ despite the presence of metrologically useful entanglement \cite{Toth:2012}. When measurements are made that can resolve the full probability distribution rather than just the mean and variance of the collective spin, as is required for resolving the full metrological potential of highly entangled quantum states \cite{Haine:2018b},  a more relevant metric for quantifying the sensitivity is the quantum Fisher information (QFI), which relates to the sensitivity via the quantum Cramer-Rao bound $\Delta \phi = 1/\sqrt{F_Q}$. For an initial pure state entering a MZ interferometer, the QFI is $F_Q = 4 \mathrm{Var}(\Jhat_y)$. However, if we allow for an additional $\Jhat_x$ rotation before the MZ interferometer, the QFI can be increased further. Specifically, we use the definition 
\begin{equation}
F_Q = 4 \mathrm{V}(\Jhat_\theta)
\end{equation}
where $\Jhat_\theta = \cos\theta \Jhat_y + \sin\theta \Jhat_z$, and $\theta$ is chosen to maximise the QFI. We note that a full probability-resolving measurement is not required if an interaction-based readout is used after the interferometer \cite{Nolan:2017b, Mirkhalaf:2018, Haine:2018b}. Figure (\ref{fig4}b) shows the QFI for both OAT and TNT for $N = 10^5$ particles. We see that TNT reaches a higher value, and achieves this maximum significantly more quickly. Specifically, TNT reaches the threshold $F_Q \approx N^2/2$, which is the QFI of the highly non-classical twin-Fock state, $\sim 40$ times faster than OAT. Alternatively, at the time when TNT reaches this threshold, the QFI is $\sim 337$ times larger than for OAT.

\section{Multi-Mode Dynamics}\label{sec4}
We now investigate the effects of multi-mode dynamics. Assuming a cigar shaped trapping potential where the two transverse dimensions ($y$ \& $z$) are much tighter than the longitudinal dimension ($x$), the dynamics in the $x$ direction is well approximated by the Heisenberg equations of motion
\begin{subequations}
\begin{align}
    i\hbar\frac{d}{dt}\psihat_a &= \left(\hat{H}_{1D} + \tilde{U}_{aa}\psihatd_a \psihat_a + \tilde{U}_{ab}\psihatd_b \psihat_b \right)\psihat_a + \frac{\hbar\Omega}{2}\psihat_b \label{psia_dot}\\
i\hbar\frac{d}{dt}\psihat_b &= \left(\hat{H}_{1D} + \tilde{U}_{ab}\psihatd_a \psihat_a + \tilde{U}_{bb}\psihatd_b \psihat_b \right)\psihat_b + \frac{\hbar\Omega}{2}\psihat_a \label{psib_dot}
\end{align}
\end{subequations}
where
\begin{align}
 \hat{H}_{1D}  &= \frac{-\hbar^2}{2m}\frac{\partial^2}{\partial x^2} + \frac{1}{2}m \omega_x^2 x^2 \, ,
\end{align}
and $\tilde{U}_{ij} = U_{ij}/A_\perp$ is the dimensionally reduced effective one dimensional interaction strength obtained by dividing the three dimensional interaction strength by a parameter characterising the transverse area of the system. Throughout this work, we use $m = 87$ amu, and $A_\perp = 10^{-10}$ m$^2$. 

\begin{figure}[htbp]
\centering{\includegraphics[width=0.8\columnwidth]{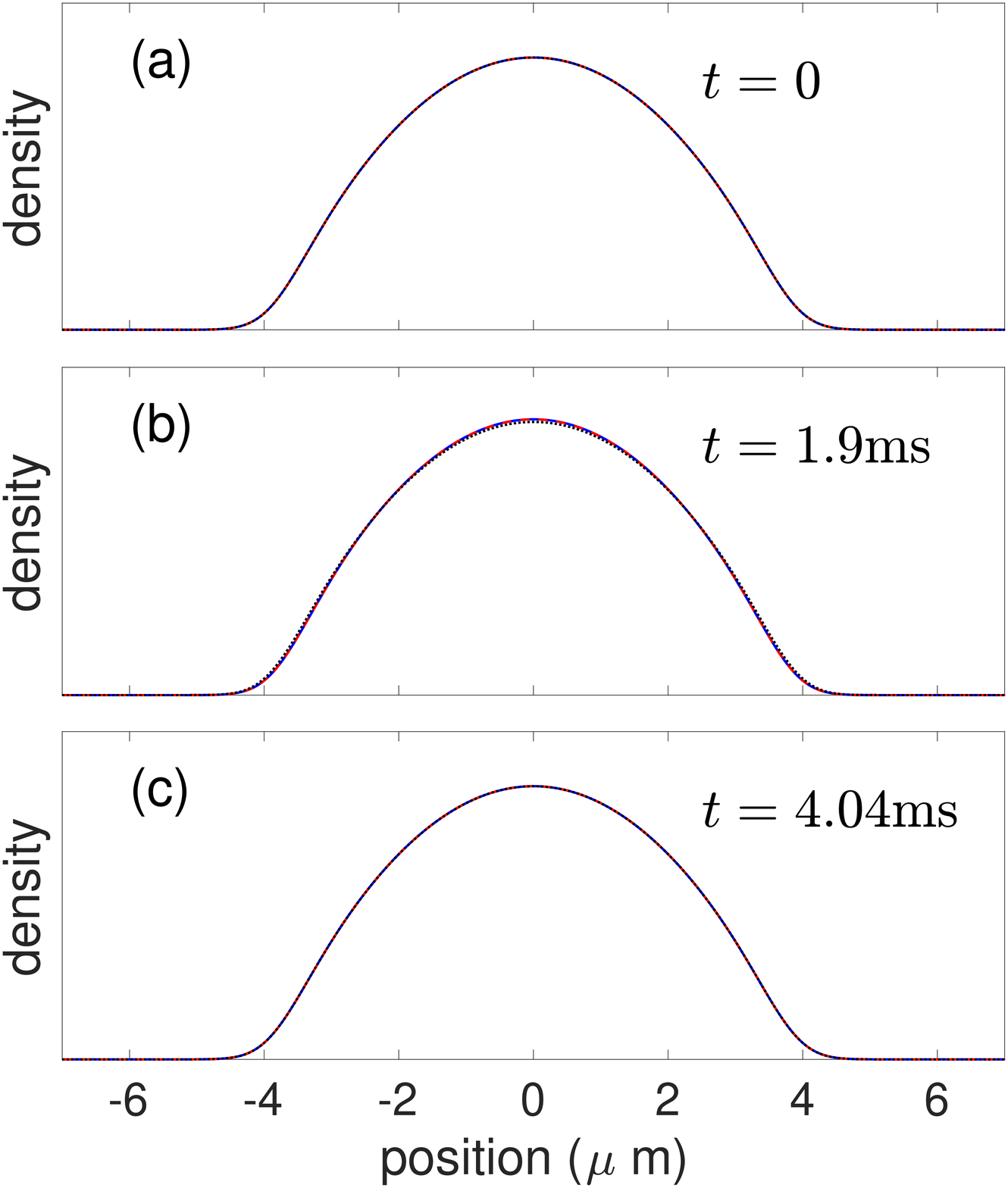}}
\caption{Evolution of the density of each component for Case I. $|\psi_a(x,t)|^2$ (blue solid line), $|\psi_b(x,t)|^2$ (red dashed line), compared to the initial state $|\psi_a(x,0)|^2 = |\psi_b(x,0)|^2 = |\psi_0(x)|^2$ (black dotted line). Both components vary only slightly from the initial condition, but remain identical to each other. }
\label{densityplot_case1}
\end{figure}

\begin{figure}[htbp]
\centering{\includegraphics[width=0.8\columnwidth]{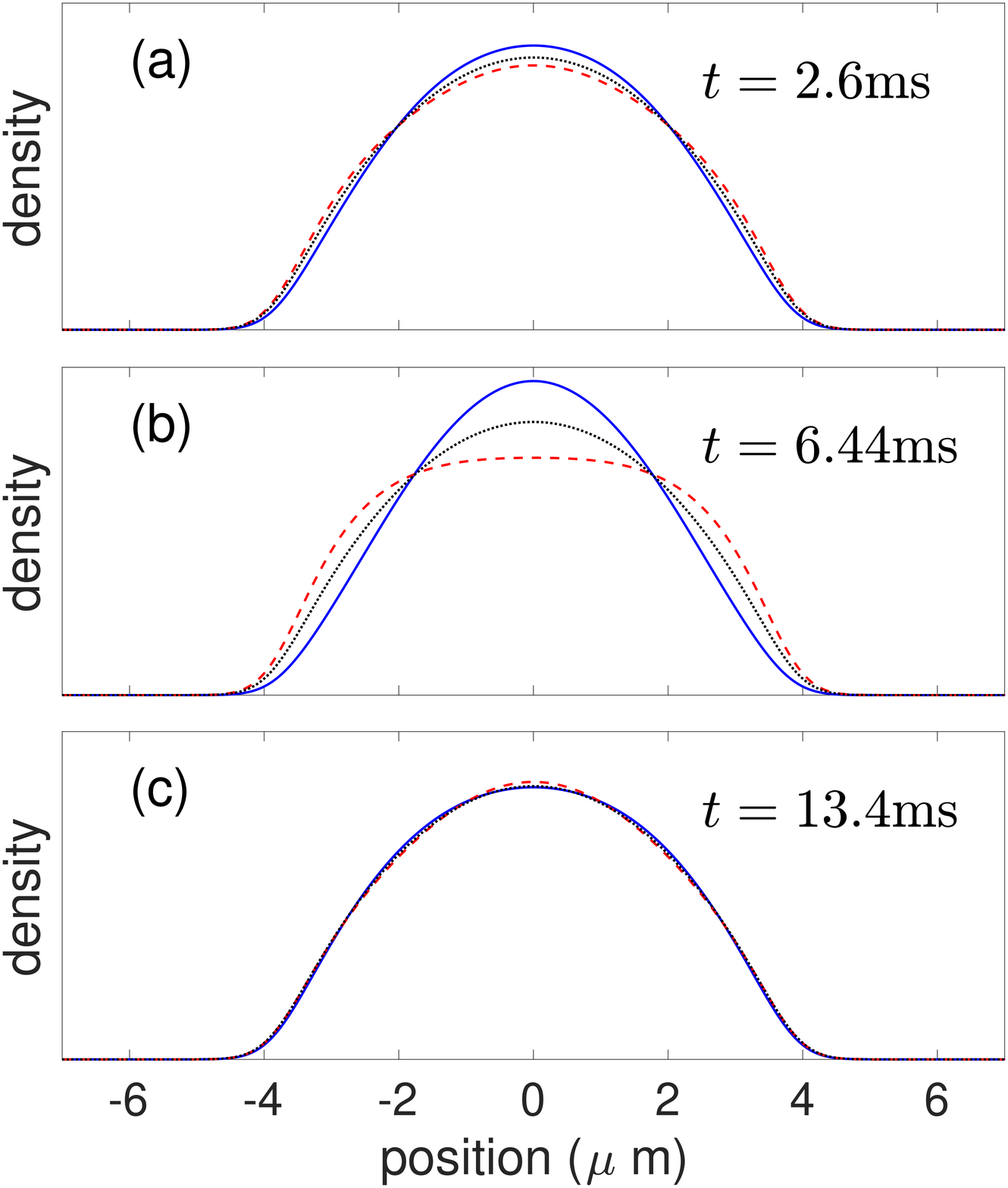}}
\caption{Evolution of the density of each component for Case II when a 50/50 beamsplitter is implemented. $|\psi_a(x,t)|^2$ (blue solid line), $|\psi_b(x,t)|^2$ (red dashed line), compared to the initial state $|\psi_a(x,0)|^2 = |\psi_b(x,0)|^2 = |\psi_0(x)|^2$ (black dotted line).}
\label{densityplot_case2(a)}
\end{figure}

\begin{figure}[htbp]
\centering{\includegraphics[width=0.8\columnwidth]{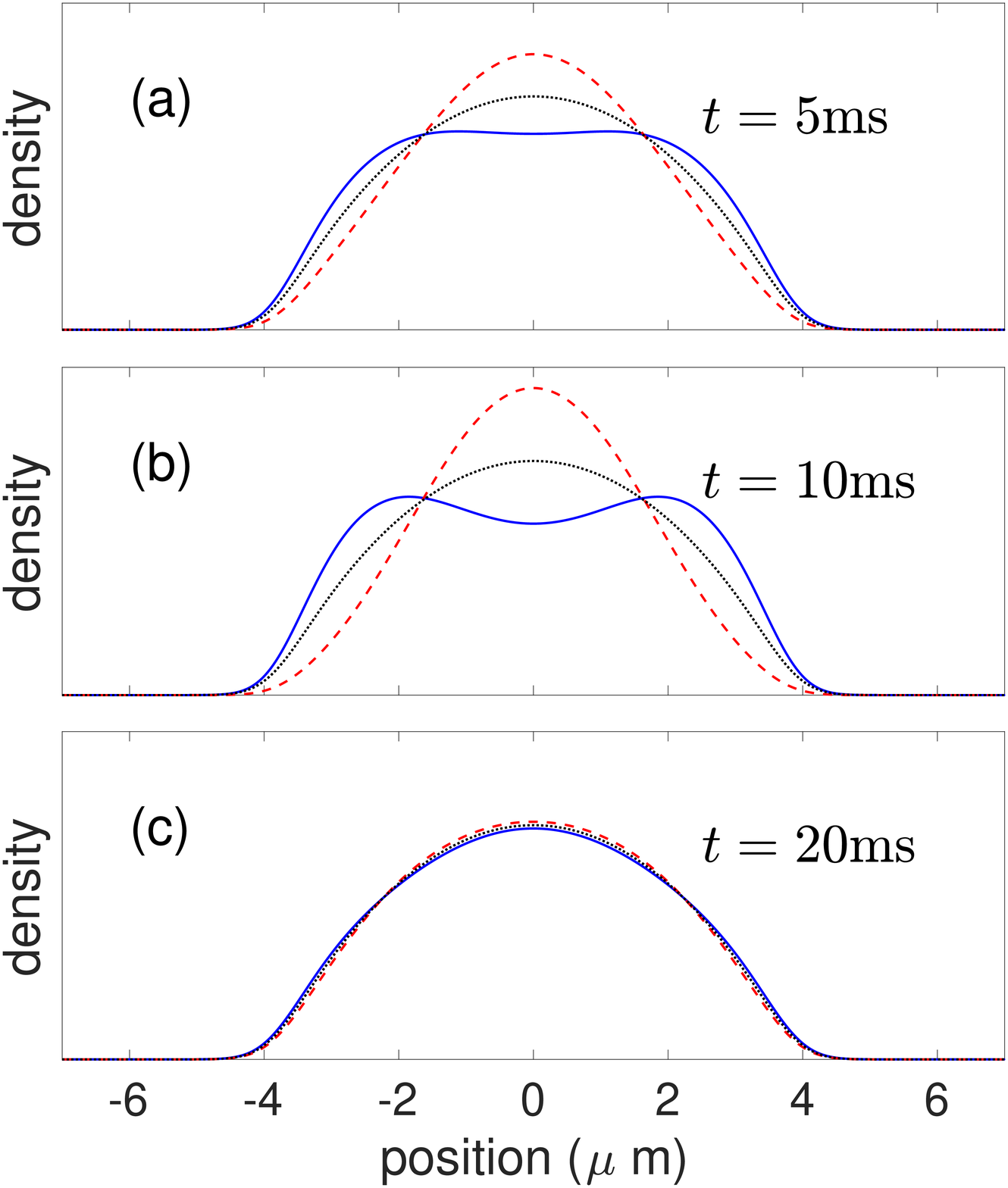}}
\caption{Evolution of the density of each component for Case III. $|\psi_a(x,t)|^2$ (blue solid line), $|\psi_b(x,t)|^2$ (red dashed line), compared to the initial state $|\psi_a(x,0)|^2 = |\psi_b(x,0)|^2 = |\psi_0(x)|^2$ (black dotted line).}
\label{densityplot_case3}
\end{figure}

We model the situation where initially all the atoms in are in state $|a\rangle$ in the ground motional state, and then coherently transfer half the population to state $|b\rangle$. Unless $U_{aa} = U_{bb} = U_{ab}$, this new state will not be the ground state, and motional dynamics will occur. As $\chi \propto U_{aa} + U_{bb} - 2U_{ab} \neq 0$, we cannot obtain entangling dynamics without also exciting motional dynamics. We consider three distinct cases that provide qualitatively different dynamics:
\begin{itemize}
    \item Case I: $U_{aa} = U_{bb} > U_{ab}$. In this case, the two components will undergo breathing oscillations, but will tend to breathe-together. Specifically, we choose $a_{aa} = a_{bb} = 100.0a_0$, $a_{ab} = 97.0 a_0$, where $a_0 =5.29\times10^{-11} $ m.
    \item Case II: $a_{bb} > a_{aa} > a_{ab}$. In this case, the components tend to separate, as one component breathes inwards while the other breathes outwards, such that the overlap of the two components is significantly decreased. However, by adjusting the relative atom numbers, a \emph{breathe-together} solution exists \cite{Sinatra:2000}. Specifically, we chose $a_{aa} = 95.0a_0$, $a_{bb} = 100.0a_0$, $a_{ab} = 90.0a_0$.
    \item Case III:  $U_{aa} > U_{ab} > U_{bb}$. The two components will tend to separate, and no breathe-together solution exists. Specifically, we choose $a_{aa} = 100.0a_0$, $a_{bb} = 95.0a_0$, $a_{ab} = 97.0a_0$. This represents the scattering parameters of the two hyperfine ground-states of $^{87}$Rb.  
\end{itemize}
To illustrate the three cases, we calculate the density distribution under the mean-field approximation by solving the Gross-Pitaevskii equation \cite{Dalfovo:1999}, obtained by making the substitution $\psihat_j(x) \rightarrow \psi_j(x)$ in equations (\ref{psia_dot},\ref{psib_dot}). Figures (\ref{densityplot_case1}, \ref{densityplot_case2(a)}, \ref{densityplot_case3}) shows the evolution of the density distribution for these three cases. In particular, note that in cases I the two components evolve together, while in case II and III, the two components spatially separate. This separation will hinder the ability to implement TNT dynamics, as the varying spatial overlap will complicate the coherent coupling required for the $\Jhat_x$ rotation. Additionally, the spin-squeezing parameter requires a high degree of overlap between the modes. \\

The Gross-Pitaeveskii equation is incapable of capturing the evolution of the quantum statistics, which is required to investigate spin-squeezing and entanglement. To investigate this effect, we simulate the dynamics of the system using the Truncated Wigner (TW) method, which has previously been used to model the dynamics of quantum gases \cite{Steel:1998, Sinatra:1995, Norrie:2006, Drummond:2017}, and unlike the GPE, can be used to model non-classical particle correlations \cite{Haine:2011, Ruostekoski:2013, Haine:2014, Szigeti:2017, Haine:2018b, Szigeti:2020}. The derivation of the TW method has been described in detail elsewhere \cite{Drummond:1993, Steel:1998, Blakie:2008}. Briefly, the equation of motion for the Wigner function of the system can be found from the von-Neumann equation by using correspondences between differential operators on the Wigner function and the original quantum operators \cite{Gardiner:2004b}. By truncating third- and higher-order derivatives (the TW approximation), a Fokker–Planck equation (FPE) is obtained. The FPE is then mapped to a set of stochastic partial differential equations for complex fields $\psi_j(x,t)$, which loosely correspond to the original field operators $\psihat_j(x, t)$, with initial conditions stochastically sampled from the appropriate Wigner distribution \cite{Blakie:2008, Olsen:2009}. The complex fields obey the partial differential equations
\begin{subequations}
\begin{align}
    i\hbar\frac{d}{dt}\psi_a &= \hat{H}_{1D}\psi_a  + \frac{\hbar\Omega}{2}\psi_b \nonumber \\
    &+ \left(\tilde{U}_{aa} \left(|\psi_a |^2 -\frac{1}{\Delta x}\right) + \tilde{U}_{ab}\left(|\psi_b |^2 -\frac{1}{\Delta x}\right)\right)\psi_a  \label{psia_TW_dot} \\
    i\hbar\frac{d}{dt}\psi_b &= \hat{H}_{1D}\psi_b  + \frac{\hbar\Omega}{2}\psi_a \nonumber \\
    &+ \left(\tilde{U}_{ab} \left(|\psi_a |^2 -\frac{1}{\Delta x}\right) + \tilde{U}_{bb}\left(|\psi_b |^2 -\frac{1}{\Delta x}\right)\right)\psi_b  \label{psib_TW_dot} 
\end{align}
\end{subequations}
where $\Delta$ is discretisation size of the spatial $x$ grid. By averaging over many trajectories with stochastically sampled initial conditions, expectation values of quantities corresponding to symmetrically ordered operators in the full quantum theory can be obtained via the correspondence $\langle \{ f(\psihat^\dag_j, \psihat_j)\}_\mathrm{sym}\rangle = \overline{f[\psi_j^*, \psi_j]}$, where `sym' denotes symmetric ordering and the overline denotes the mean over many stochastic trajectories. The initial conditions for the simulations are chosen as $\psi_a(x,0) = \Psi_0(x) + \eta_a(x)$, $\psi_b(x,0) = \eta_b(x)$, where $\Psi_0(x)$ is the ground state of the single-component time-independent Gross-Pitaevski equation, and $\eta_j(\xi)$ are complex Gaussian noises satisfying $\overline{\eta^*_i(x_n)\eta_j(x_m)} = \frac{1}{2}\delta_{m,n}\delta_{i,j}/\Delta$, for spatial grid points $x_m$ and $x_n$. At $t=0$, the $\pi/2$ beam splitting pulse which initiates the dynamics is implemented via the transformation $\psi_a(x) = \frac{1}{\sqrt{2}}\left(\psi_a(x,0) + \psi_b(x,0)\right)$, $\psi_a(x) = \frac{1}{\sqrt{2}}\left(\psi_b(x,0) - \psi_a(x,0)\right)$.

\subsection{Case: I}
\begin{figure}
\centering{\includegraphics[width=\columnwidth]{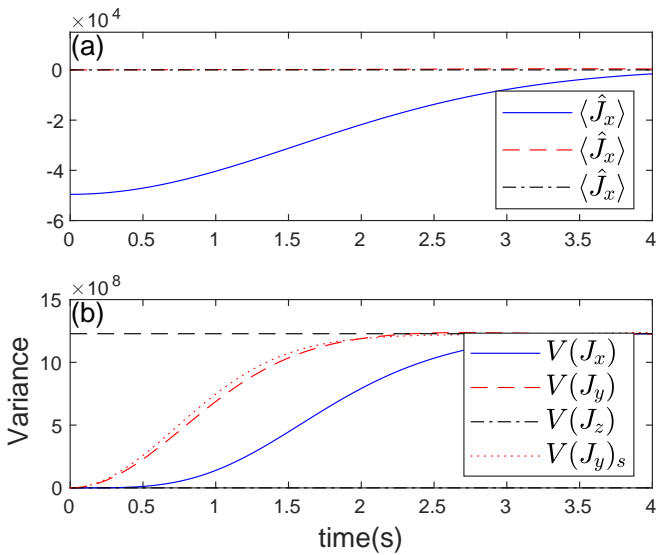}}
\caption{Comparison between the multi-mode dynamics and ideal single-mode behaviour for Case I, when only OAT dynamics is implemented. When we choose $\chi = 2.2 \times 10^{-3}$ s$^{-1}$, we have excellent agreement in the variance of the pseudospins. $V(J_y)$ as calculated from the single-mode model is indicated via the red dotted line.}
\label{MMOAT_case1}
\end{figure}

In order to implement TNT dynamics, the rotation parameter $\Omega$ must be set to the optimum value $\Omega = \chi N/2$. While this is simple to do in the single-mode model, in the multimode model the effective $\chi$ depends on the time-dependent density distributions. Furthermore, it has previously been shown that when strong multimode dynamics are present, estimates of $\chi$ based on \eq{chi_est} are poor \cite{Haine:2014}. We estimate the effective $\chi$ by first setting $\Omega=0$ and calculating the variance of the three pseudo-spin operators, and comparing to ideal single-mode OAT dynamics for the same number of atoms. The single-mode dynamics resulting from \eq{Ham_OAT} is also calculated via the TW method \footnote{The quantum dynamics for the single-mode model is obtained using the TW method by solving the ODEs $i\dot{\alpha} = \frac{1}{2}\chi(|\alpha|^2 - |\beta|^2)\alpha + \Omega \beta $, $i\dot{\beta} = -\frac{1}{2}\chi(|\alpha|^2 - |\beta|^2)\beta + \Omega \alpha$, where $\alpha$ and $\beta$ are the stochastic variables corresponding to $\ahat_0$ and $\bhat_0$ respectively.}. Figure (\ref{MMOAT_case1}) shows a comparison between the multi-mode dynamics and single-mode dynamics, with the parameter $\chi$ in the single-mode mode adjusted to provide the best match to the multimode dynamics.  We use this as our estimate of $\chi$ when choosing a value of $\Omega$ when implementing TNT dynamics in the multi-mode model. 

\begin{figure}
\centering{\includegraphics[width=\columnwidth]{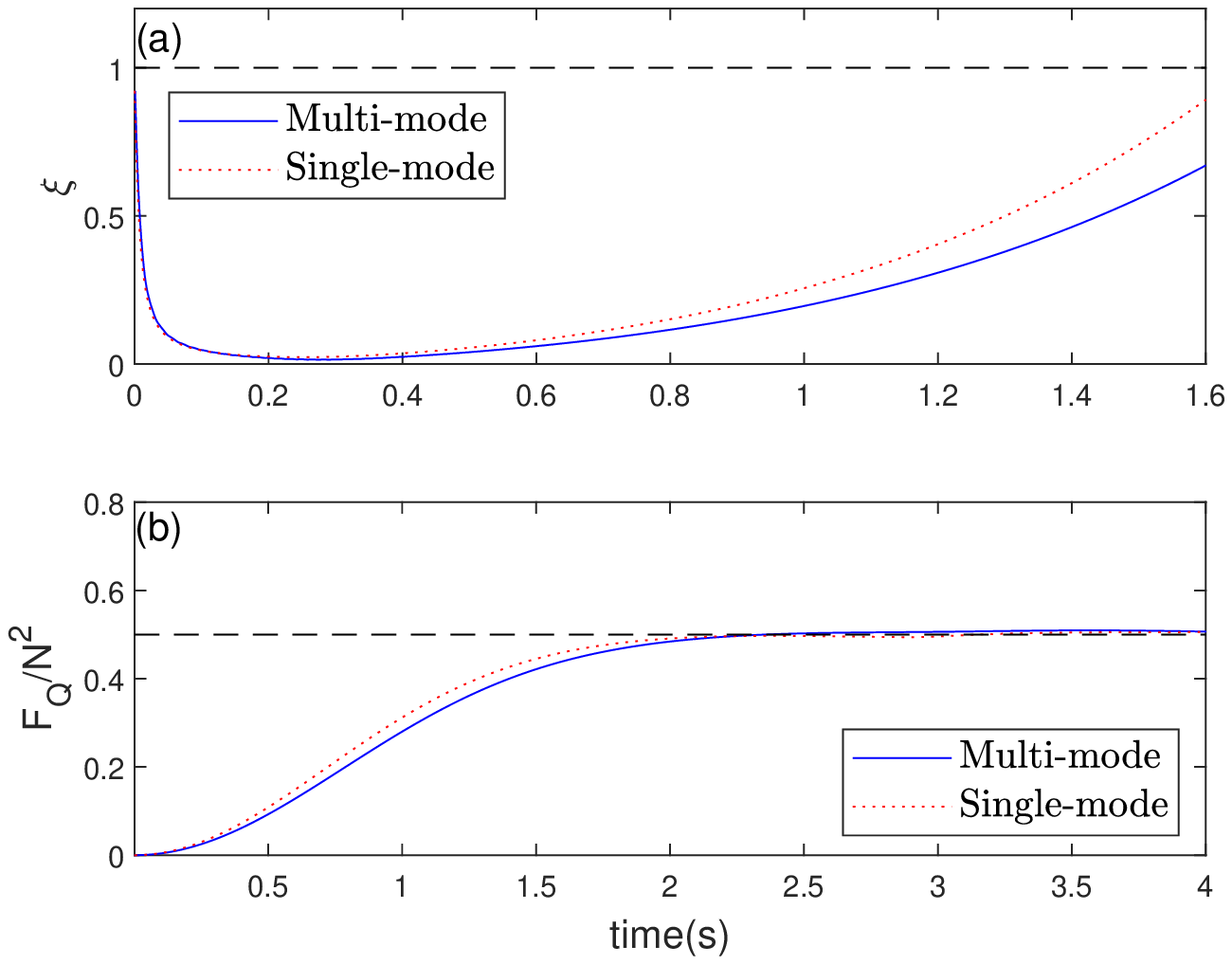}}
\caption{Comparison between multi-mode OAT dynamics and ideal OAT single-mode behaviour for Case I, for the spin-squeezing parameter (a) and QFI (b).}
\label{fig:case1_xi}
\end{figure}

\begin{figure}
\centering{\includegraphics[width=\columnwidth]{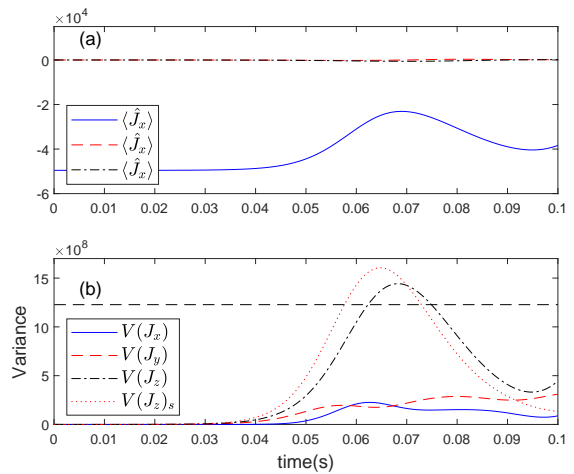}}
\caption{Comparison between the multi-mode dynamics and ideal single-mode behavior when TNT dynamics is implemented. The value of $\Omega$ is chosen as $\Omega = \chi N/2$, where the value $\chi = 2.2 \times 10^{-3}$ s$^{-1}$, was used. $V(J_z)$ as calculated from the single-mode model is indicated via the red dotted line. }
\label{MMTNT_case1}
\end{figure}

\begin{figure}
\centering{\includegraphics[width=\columnwidth]{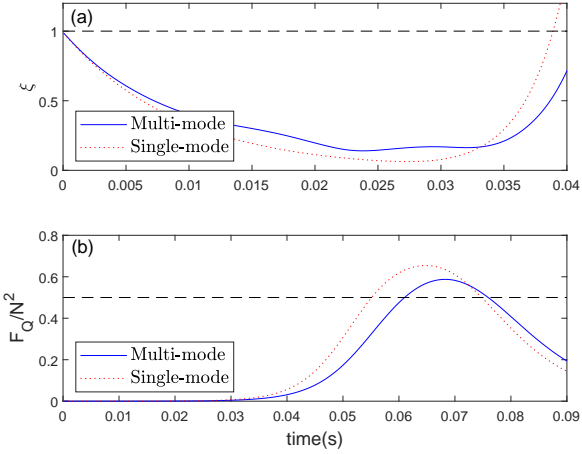}}
\caption{Comparison between multi-mode TNT dynamics and ideal TNT single-mode behaviour for Case I, for the spin-squeezing parameter (a) and QFI (b)}. 
\label{fig:MMTNT_xi}
\end{figure}

We see here that setting $\chi = 2.2\times 10^{-3}$ s$^{-1}$ provides excellent agreement between the single-mode model and the multimode model for both the pseudospin variances (fig. \ref{MMOAT_case1}) spin-squeezing parameter, and QFI (fig.\ref{fig:case1_xi}). Using this value in our choice of $\Omega = N \chi/2$ should therefore result in TNT dynamics. Figure \ref{MMTNT_case1} shows the multimode dynamics compared to the single mode dynamics. We see that there is good agreement between the single-mode and multi-mode models, which is unsurprising given that the two components evolve identically, ensuring that the overlap between the two components remains constant, and that the relative phase is the same at each point in space. This ensures the coupling term results in the pure $J_x$ rotation required for TNT dynamics. In this parameter regime, the spin squeezing parameter and QFI also display excellent agreement with the ideal single-mode dynamics (figure \ref{fig:MMTNT_xi}). Importantly, we see that development of entanglement (characterised by large quantum Fisher information) occurs much more rapidly than with OAT dynamics.

\subsection{Case II}
\begin{figure}[htbp]
\centering{\includegraphics[width=0.8\columnwidth]{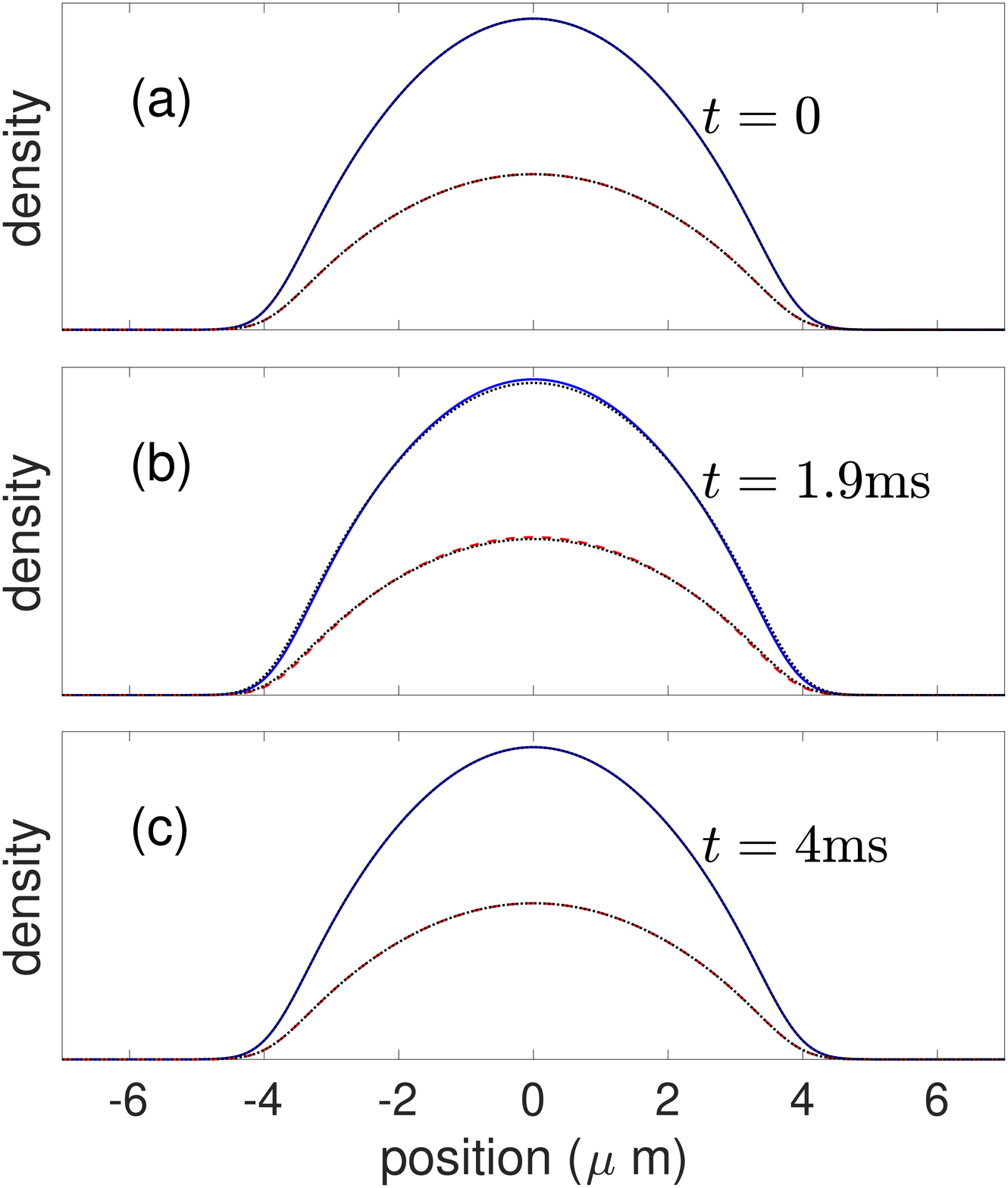}}
\caption{Evolution of the density of each component for Case II when an asymmetric beamspitter is implemented, in order to satisfy the breathe-together solution. $|\psi_a(x,t)|^2$ (blue solid line), $|\psi_b(x,t)|^2$ (red dashed line), compared to the initial conditions $|\psi_a(x,0)|^2$ and $|\psi_b(x,0)|^2$ (black dotted lines). Both components deviate only slightly from their initial conditions. }
\label{densityplot_case2(b)}
\end{figure}

\begin{figure}[htbp]
\centering{\includegraphics[width=\columnwidth]{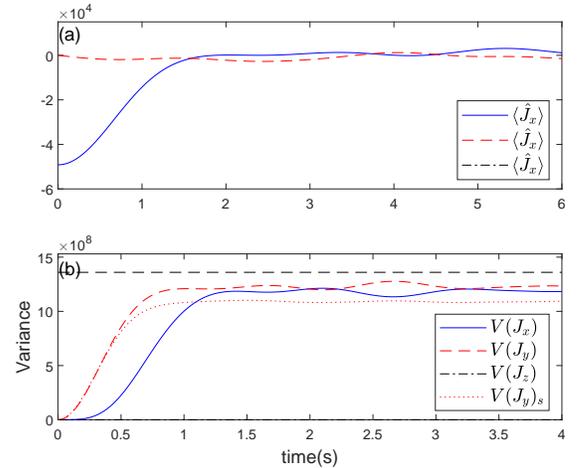}}
\caption{Comparison between multi-mode OAT dynamics and ideal OAT single-mode behaviour for Case II with the breathe-together initial condition. For the single-mode simulation, $\chi = 5.25\times 10^{-3}$ Hz was found to have the best agreement with the multimode results. $V(J_y)$ as calculated from the single-mode model is indicated via the red dotted line. }
\label{fig:caseIIoatspin}
\end{figure}

When $U_{bb} \neq U_{aa}$, the two components will evolve differently, effecting both the spatial overlap, and the relative phase. As before, we first simulate the system with $\Omega = 0$ to investigate the agreement between the single-mode and multi-mode systems for OAT dynamics. However, as can be seen in figure (\ref{densityplot_case2(a)}), the two components begin to separate, which will inhibit the performance of the TNT dynamics. In order to prevent this, we exploit the \emph{breathe-together} solution \cite{Li:2008}. This is achieved by replacing the initial 50/50 beam-splitter with an asymmetric beamsplitter such that each component experiences the same interaction strength. Specifically, this is achieved by choosing the beam-splitter angle such that the ratio of population in each component, $N_a$ and $N_b$, satisfy
\begin{equation}
\frac{N_a}{N_b} = \frac{U_{bb} - U_{ab}}{U_{aa}-U_{ab}} \,.
\end{equation}
As can be seen from figure \ref{densityplot_case2(b)}, this choice of initial condition results in the motional dynamics being frozen out. The dynamics of the quantum statistics, however, is shown in figure \ref{fig:caseIIoatspin}, as well as the dynamics for the equivalent single-mode system. The different initial conditions result in reduced final spin variances for both the multi-mode and ideal single-mode dynamics. By comparing both cases, we can infer that the effective interaction parameter $\chi$ is $\chi=5.25\times 10^{-3}$ Hz. Again, we use this value to choose the appropriate value of $\Omega$ for implementation of TNT dynamics. However, as the breathe-together solution is asymmetric in population, the optimum value of $\Omega$ will be slightly different. We account for this by trialling a range of values, as shown in figure \ref{fig:caseIItntspin}. We find that the most effective entangling dynamics (that is, the dynamics that leads to the largest spin-variance) is when $\Omega =0.85 \chi N/2$. 

\begin{figure}[htbp]
\centering{\includegraphics[width=\columnwidth]{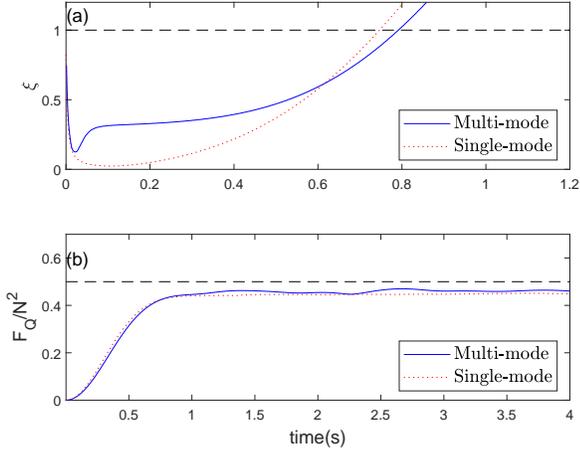}}
\caption{Comparison between multi-mode OAT dynamics and ideal OAT single-mode behaviour for Case II with the breathe-together solution, for the spin-squeezing parameter (a) and QFI (b).}
\label{fig:caseIIoatxi}
\end{figure}

\begin{figure}[htbp]
\centering{\includegraphics[width=\columnwidth]{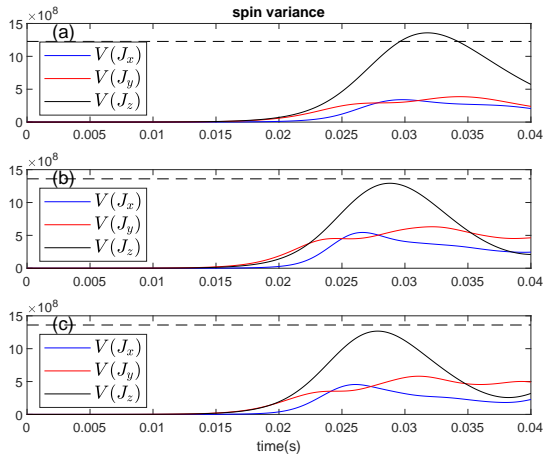}}
\caption{Comparison between single-mode TNT dynamics (a) and multimode TNT dynamics ((b) and (c)). The initial condition was chosen to satisfy the breathe-together solution. In (a) and (b), the optimal rotation rate ($\Omega = 0.85 \chi N/2$) was used, which gives slightly better performance than the usual TNT solution ($\Omega = \chi N/2$), shown in (c).}
\label{fig:caseIItntspin}
\end{figure}

\begin{figure}[htbp]
\centering{\includegraphics[width=\columnwidth]{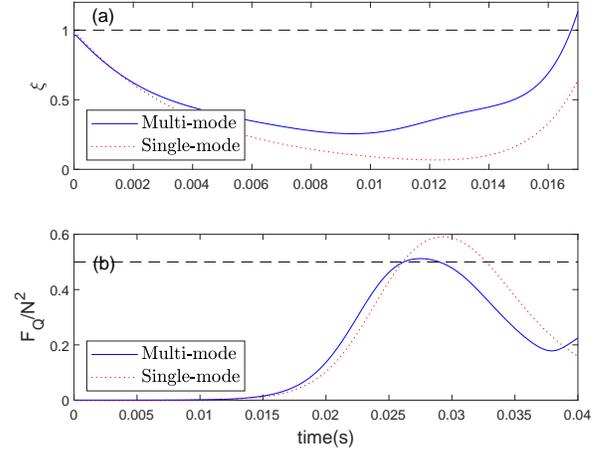}}
\caption{Comparison between multi-mode TNT dynamics and ideal TNT single-mode behaviour for Case II with the breathe-together solution, for the spin-squeezing parameter (a) and QFI (b). $\Omega = 0.85 \chi N/2$ was used for both cases.}
\label{fig:caseIItntxi2}
\end{figure}

In figure \ref{fig:caseIItntxi2} we see good agreement between the ideal single-mode and multi-mode TNT dynamics  for the spin-squeezing parameter and QFI. Importantly, the entangling dynamics occurs $\sim 40$ times faster than for OAT alone.

\subsection{Case III}
\begin{figure}[htbp]
\centering{\includegraphics[width=\columnwidth]{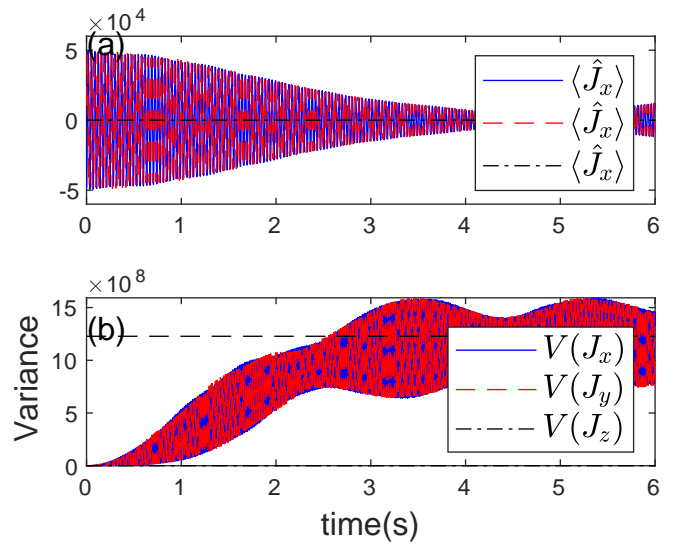}}
\caption{Multimode OAT dynamics behaviour for Case III when 50/50 beamspliter is implemented. (a) the expectation values of the spin operators, and (b) the variances of the spin operators}
\label{fig:caseIIIoatspin}
\end{figure}

\begin{figure}[htbp]
\centering{\includegraphics[width=\columnwidth]{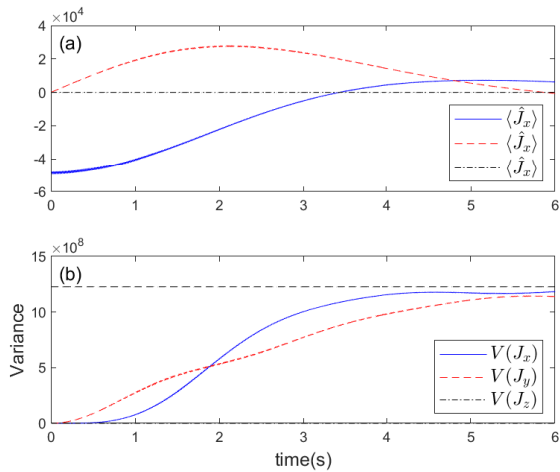}}
\caption{Multimode OAT dynamics behaviour for Case III when a rotation about the $\hat{J}_z$ axis is added to compensate for the spin-precession. (a) the expectation values of the spin operators, and (b) the variances of the spin operators.}
\label{fig:caseIIIOATspin}
\end{figure}

When $U_{aa} > U_{ab} > U_{bb}$, there is no breathe-together solution. As before, we first simulate the system with $\Omega = 0$ to investigate the behaviour in a multimode system for OAT dynamics. However, as can be seen in figure (7), the two components begin to separate, which will inhibit the performance of the TNT dynamics. Additionally, as can be seen in figure \ref{fig:caseIIIoatspin}, the asymmetry in scattering lengths results in an additional rotation of the collective spin around the $J_z$ axis, which will further inhibit TNT dynamics, as we require a coherent rotation around the collective spin direction. We can correct for this term by adding an additional rotation of angular frequency $\omega_r$, either by adjusting the detuning between the two levels, or by dynamically rotating the relevant rotation axis for our TNT dynamics. Figure \ref{fig:caseIIIOATspin} shows the spin dynamics of the system under OAT dynamics with this additional correction. While not mimicking the single mode OAT dynamics perfectly, it displays qualitatively similar behaviour. 

\begin{figure}[htbp]
\centering{\includegraphics[width=\columnwidth]{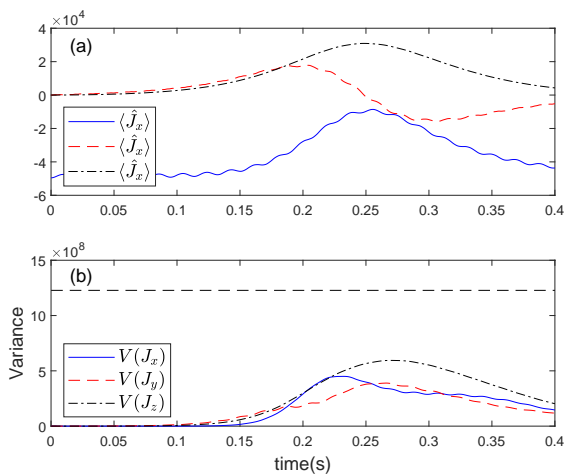}}
\caption{Multimode TNT dynamics behaviour for Case III, for the spin variance (a) and three mean value of angular operators (b).}
\label{fig:caseIIItntspin}
\end{figure}

We have modelled TNT dynamics with this additional rotation term (figure \ref{fig:caseIIItntspin}). We see that, while the timescale of the entangling dynamics is significantly faster than OAT alone, it does not behave as well as ideal TNT dynamics. The reason for this is that while the additional rotation partially corrects for the drifting phase difference between the two components, the multimode dynamics ensures that there is a dynamic and spatially-varying phase difference, so this cancellation is imperfect. This effect has an even more pronounced effect on the spin-squeezing and QFI (figure \ref{fig:caseIIItntxi2}); there is little improvement in either the rate or depth of spin-squeezing achievable by implementing TNT dynamics. 

\begin{figure}[htbp]
\centering{\includegraphics[width=\columnwidth]{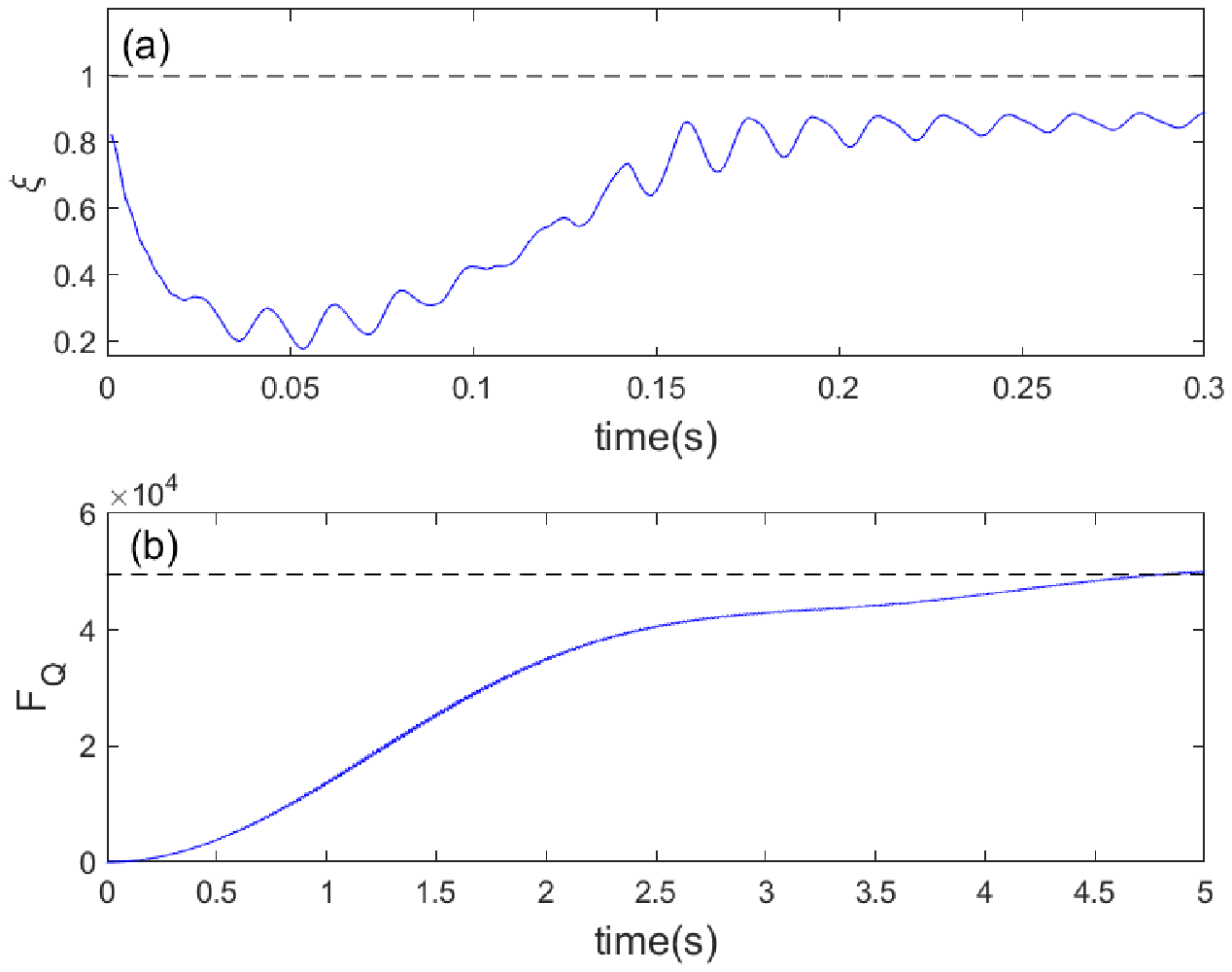}}
\caption{Multimode OAT dynamics behaviour for Case III, for the spin squeeze parameter (a) and QFI (b).}
\label{fig:caseIIIoatxi}
\end{figure}
\begin{figure}[htbp]
\centering{\includegraphics[width=\columnwidth]{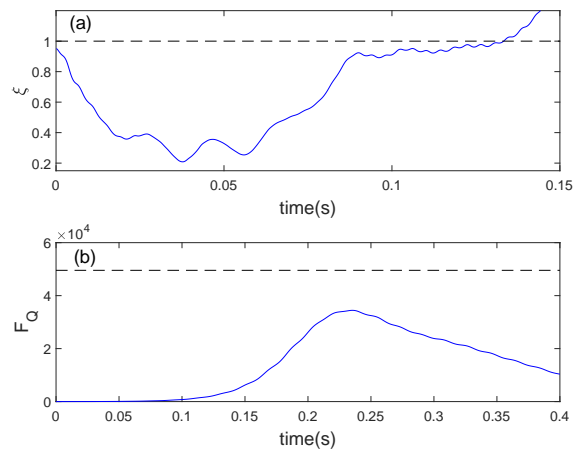}}
\caption{Multimode TNT dynamics behaviour for Case III, for the spin squeeze parameter (a) and QFI (b).}
\label{fig:caseIIItntxi2}
\end{figure}

\section{Discussion}\label{sec5}
Our modelling indicates that in case I and case II systems, TNT can be used to significantly speed up spin-squeezing and entangling dynamics even in regimes where significant multi-mode dynamics is present. Importantly, in case II systems, use of the breathe-together solution can be used to ensure strong mode-overlap between the two components. In both of these cases, the spin-squeezing dynamics of the full system is well approximated by an effective two-mode model. 

However, in case III systems, the implementation of TNT dynamics provides little-to-no benefit over conventional OAT dynamics. We attribute this to a spatially varying phase-profile of the two components, which results in a variation in the rotation axis of the effective TNT rotation.  

Our results indicate that in some regimes, spin-squeezing and entanglement generation via atomic self-interaction is achievable in BECs with large numbers of atoms, even in the presence of multi-mode dynamics, and that TNT dynamics can be used to decrease the state-preparation time required in order to achieve this squeezing. Alternatively, faster entangling dynamics can result in better spin-squeezing in cases where the state-preparation time is limited \cite{Haine:2020, Szigeti:2020, Szigeti:2021}.

\section{ACKNOWLEDGMENTS}
This research was undertaken with the assistance of re-sources and services from the National Computational Infrastructure (NCI), which is supported by the Australian Government. The Australian National University is situated on land traditionally owned by the Ngunnawal people.

\bibliography{tnt_bib.bib}

\end{document}